\documentclass{emulateapj}
\usepackage[dvips]{color}
\usepackage{amssymb, amsmath, amsbsy, epsfig, epsf, slashbox}
\usepackage{graphicx}
\usepackage{wrapfig}
\usepackage{color}

\usepackage[T1]{fontenc}
\usepackage{float}
\usepackage{amsthm}
\usepackage{algorithm}
\usepackage{algorithmic}
\usepackage{epstopdf}

\newcommand{\lbol}{\ensuremath{L\mathrm{_{bol}}}}

\newcommand{\kms}{\ensuremath{\mathrm{km~s^{-1}}}}

\newcommand{\ha}{{\rm H\ensuremath{\alpha}}}
\newcommand{\hb}{{\rm H\ensuremath{\beta}}}

\newcommand{\hii}{H\,{\footnotesize II}}

\newcommand{\oiil}{[O\,{\footnotesize II}] $\lambda$3727}
\newcommand{\neiiil}{[Ne\,{\footnotesize III}] $\lambda$3870 }
\newcommand{\oiiil}{[O\,{\footnotesize III}] $\lambda$5007}
\newcommand{\oil}{{\rm [O\,{\footnotesize I}]}$\lambda$6300}
\newcommand{\niir}{[N\,{\footnotesize II}] $\lambda$6583}
\newcommand{\niil}{[N\,{\footnotesize II}] $\lambda$6548}
\newcommand{\siil}{[S\,{\footnotesize II}] $\lambda \lambda$6717, 6731}

\newcommand{\oiii}{[O\,{\footnotesize III}]}

\newcommand{\oii}{[O\,{\footnotesize II}]}

\newcommand{\neiii}{[Ne\,{\footnotesize III}]}
\newcommand{\oi}{{\rm [O\,{\footnotesize I}]}}
\newcommand{\nii}{[N\,{\footnotesize II}]}
\newcommand{\sii}{[S\,{\footnotesize II}]}

\def\lax{{$\mathrel{\hbox{\rlap{\hbox{\lower4pt\hbox{$\sim$}}}\hbox{$<$}}}$}}
\def\gax{{$\mathrel{\hbox{\rlap{\hbox{\lower4pt\hbox{$\sim$}}}\hbox{$>$}}}$}}

\def\kms{\hbox{km$\,$s$^{-1}$}}

\begin{document}


\title{ A New Diagnostic Diagram of Ionization Source for High Redshift Emission Line Galaxies}

\author{
Kai~Zhang\altaffilmark{1,2}, Lei~Hao\altaffilmark{1} }

\altaffiltext{1}{Key Laboratory for Research in Galaxies and Cosmology, Shanghai
Astronomical Observatory, Chinese Academy of Sciences, 80 Nandan
Road, Shanghai 200030, China}

\altaffiltext{2}{Department of Physics and Astronomy, University of Kentucky, 505 Rose Street, Lexington, KY 40506, USA}

\email{Correspondence should be addressed to Lei Hao: haol@shao.ac.cn}
\shorttitle{The KEx diagram}
\shortauthors{Zhang et al.}

\begin{abstract}
We propose a new diagram, the Kinematic-Excitation diagram (KEx diagram), which uses the \oiii/\hb\ line ratio and the \oiiil\
emission line width ($\sigma_{\oiii}$) to diagnose the ionization source and physical properties of the 
Active Galactic Nuclei (AGNs) and the star-forming galaxies (SFGs).
The KEx diagram is a suitable tool to classify emission-line galaxies (ELGs) at intermediate redshift because it uses $only$ the \oiiil\ and \hb\ emission lines. 
We use the SDSS DR7 main galaxy sample and the Baldwin$-$Phillips$-$Terlevich (BPT) diagnostic to calibrate the diagram at low redshift.
We find that the diagram can be divided into 3 regions: one occupied mainly
by the pure AGNs (KEx-AGN region), one dominated by composite galaxies
(KEx-composite region), and one contains mostly SFGs (KEx-SFG region).
AGNs are separated from SFGs in this diagram mainly because they
preferentially reside in luminous and massive galaxies and have high \oiii/\hb.  The separation of AGN from star-forming galaxies is even
cleaner thanks to the additional 0.15/0.12~dex offset in $\sigma_{\oiii}$ at fixed luminosity/stellar mass.

We apply the KEx diagram to 7,866 galaxies at 0.3 $<$ z $<$ 1 in the DEEP2 Galaxy Redshift Survey,
and compare it to an independent X-ray classification scheme using $Chandra$ observation. X-ray AGNs are mostly 
located in the KEx-AGN region while X-ray SFGs are mostly located in the KEx-SFG region. 
Almost all of Type1 AGNs lie in the KEx-AGN region. These confirm the reliability of this classification diagram for emission line galaxies at intermediate redshift.
At z$\sim$2,  the demarcation line between star-forming galaxies and AGNs should shift 0.3~dex higher in $\sigma_{\oiii}$ to account for evolution. 

%

\end{abstract}


\section{Introduction}
The diagnostic diagrams are major tools to understand the nature of galaxies. 
They are crucial in evaluating the galaxy evolution scenarios such as the cosmic accretion and star-formation histories.
The most widely-used diagram is the BPT (Baldwin, Philips \& Terlevich 1981)
or VO87 diagram (Veilleux \& Osterbrock 1987).
The advent of spectroscopic sky surveys like the SDSS
and the photoionization models (Ferland et al. 1998) make the classification observationally constrained
and theoretically understood. Kewley et al. (2001) used a variety of \hii\ region photoionization
models to give a theoretical star-forming galaxy boundary on the BPT diagram. The sources above this
line are unlikely to be ionized by stars. Kauffmann et al. (2003) used the SDSS main galaxies
sample to map their detailed distribution on the BPT diagram, and proposed that the right branch
of the seagull shape distribution are all AGNs. The sources lie between the two dividing
lines are called composite galaxies because their gas may be ionized by
AGN and SF at the same time. Kewley et al. (2006) further proposed low ionization line criteria
for separating Seyfert2s and LINERs. Other refinement of the classification are
proposed by many authors (e.g., Stasi\'{n}ska et al. 2006; Cid Fernandes et al. 2010, 2011).

Our understanding of the BPT diagram is very comprehensive. The y-axis of the BPT diagram reflects mainly the ionization parameter while the x-axis is mostly determined by the metallicity
(Storchi-Bergmann, Calzetti, \& Kinney 1994, Raimann et al. 2000, Denicol\'{o} et
al. 2002; Pettini \& Pagel 2004; Stasi\'{n}ska et al. 2006; Groves et al. 2004a,b;
Groves et al. 2006; Kewley \& Ellison 2008).
The distinguishing power of the BPT diagram relies on the fact that
the AGN radiation is harder than star-forming galaxies at similar stellar mass, AGNs have higher ionization parameters, 
and AGNs reside exclusively in
massive metal-rich galaxies (Kauffmann et al. 2003, Groves et al. 2006).

The BPT diagram, however, suffers a few limitations.
It needs at least 4 lines (\oiiil, \hb, \niir, \ha) to make a classification.
When the strength of these 4 lines are similar, the more signal-to-noise ratio cuts
imposed, the more sources are missed. 
The \niir\ and \ha\ emission line will shift out of
the optical wavelength range when the redshift is greater than 0.4, making the classification
diagram futile for higher redshift sources with optical spectrum only.

With spectroscopic sky surveys pushing to higher redshift and
fainter luminosities, the need for a good classification diagram for emission line galaxies (ELGs) at higher redshift  is compelling.
Some efforts have been made to develop diagnostic
diagrams with spectral features in narrower wavelength range.
Tresse et al. (1996) and Rola et al. (1997)
proposed to use the of EW(\oiil) (equivalent width of \oiil), EW(\oiiil) and EW(\hb) for galaxy classification.
Stasi\'{n}ska et al. (2006) studied using
\oiil\ for galaxy classification, and proposed a method that uses 4000\AA\ break: $D_n(4000)$, EW(\oiil), and
EW(\neiiil) (DEW diagram) to select pure AGNs with z$<$1.3 using only the optical spectra.
Trouille et al. (2011) proposed to use $g-z$, \neiii, and \oii\ to clearly separate AGNs
from star-forming galaxies at intermediate redshift.
A fruitful way to push to high redshift is to retain the \oiii/\hb\ while replacing
the \nii/\ha\ with other quantities like H band absolute magnitude (Weiner et al. 2006),
\oii/\hb\ (Lamareille 2010), U-B color (Yan et al. 2011), or stellar mass (Juneau et al. 2011, 2013,
Mass-Excitation diagnostic, MEx). Marocco et al. (2011) also used $D_n(4000)$ vs \oiii/\hb\ for high-z 
galaxies classification. These methods take advantage of the fact that AGNs
reside in massive, red galaxies in local universe, and are in general efficient in
separating pure AGNs and star-forming galaxies. The composite galaxies, however, are mixed with
Seyfert2s or star-forming galaxies on these diagrams. 

The emission line velocity dispersion ($\sigma$) may trace the kinematics of different components in
AGNs and star-forming galaxies. \oiii\ in AGN comes from
the narrow line region, which better traces the bulge
kinematics (e.g. Ho 2009). \oiii\ in star-forming galaxies mainly comes from the
\hii\ regions, which locate mainly in the disk. The kinematic of the
bulge/disk is expected to be different. Catinella et al. (2010) showed that the velocity dispersion is different for bulge and disk dominated galaxies at given baryonic mass.
Besides, emission lines of AGN have extra broadening due to outflows (Greene \& Ho 2005; Zhang et al. 2011).
In principle, we could use the width of the narrow emission lines as a proxy of
the influence of bulge potential for AGNs/star-forming galaxies classification.
Following the idea of simplifying the BPT diagram as introduced in last paragraph (Weiner et al. 2006; 
Lamareille 2010; Yan et al. 2011; Juneau et al. 2011, 2013), and  the idea of different kinematics of 
AGN and star-forming galaxies, we propose to replace the \nii/\ha\ in the BPT
diagram with $\sigma_{\oiii}$ (or $\sigma_{gas}$ in general) to
separate AGNs from star-forming galaxies at high redshift. In
  Section~2, we give descriptions of the data we use. In Section~3, a
  new diagnostic diagram: the Kinematics-Excitation Diagram (KEx
  diagram) is proposed, and we explain why it works and calibrate it
  at z$<$0.3. Section~4 gives the calibration of the KEx diagram at
  0.3$<$z$<$1, and Section~5 gives the calibration at
  z$\sim$2. Discussion is given in Section~6.
We use a cosmology with $H_{\rm 0}$ = 70 km\,s$^{-1}$\,Mpc$^{-1}$, $\Omega_{\rm m}$ = 0.3, and
$\Omega_{\rm \Lambda}$ = 0.7 throughout this paper.

\section{Sample and Measurements}
\subsection{Low-redshift data}
We start from the main galaxy sample (Strauss et al. 2002) of the Sloan Digitial Sky Survey
Data Release 7 (Abazajian et al. 2009). The sample  is complete in r-band
Petrosian magnitude between 15 and 17.77 over 9380 $deg^2$. We limit the redshift range to z $<$ 0.33,
and there are 835,410 spectroscopic galaxies.
To properly measure the emission lines, we use the scheme developed in Hao et al. (2005) to
subtract the stellar absorptions.They used several hundreds SDSS low redshift pure absorption 
galaxies to construct the PCA eignspectra, and used the first 8 eignspectra to fit the continuum. 
An A-star template is added to represent young stellar population. A power law is also added when fitting AGN spectra. 
The continuum-subtracted line emissions are left for refined line fitting.
We use 1-gaussian function to fit the \oiil, \oil, \hb, \oiiil, \niil, \ha, \niir, and \siil\ (Hereafter
\oii, \oi, \hb, \oiii, \niil, \ha, \niir, and \sii)  respectively.
The $\sigma$ (line width, 1/2.35 Full Width at Half Maximum in \kms) of the gaussian
profile for \oiiil\ is denoted as $\sigma_{\oiii}$. The line ratio
of \niir/\niil\ is fixed to 3 and their profile and center are tied to be the same.
In addition, we fit \ha\ and \hb\ a second time adding one broad gaussian to account
for possible \ha\ and \hb\ broad lines. The lower limit of $\sigma$ of the broad component is 400\kms.  
The typical FWHM of \ha\ broad component of Type1 AGNs is larger than
1200\kms\ (Hao et al. 2005). We regard the broad \ha\ component to be prominent if a F-test suggests
the improvement is significant at 3$\sigma$ level.
The intrinsic velocity dispersion $\sigma_{int}$ of the emission lines is obtained by
subtracting the instrument resolution of $\sim56km/s$ using $\sigma_{int}^2=\sigma_{obs}^2-\sigma_{Instrument}^2$.
The errors of the $\sigma$ and line strength are obtained by the MPFIT
package which only includes the fitting errors (Markwardt 2009).
We perform a simple test to check how well we can measure the line width. We add a gaussian to a continuum 
with a given equivalent width (EW). The $\sigma$ of the gaussian is the combination of emission line intrinsic width 
$10^{1.8}=63.1\kms$ (typical star-forming galaxy) and the instrumental resolution of 56\kms. Random errors are added 
according to $S/N$=3, 5, 7, 10. We fit the emission line using the MPFIT package, and measure the error of $\sigma$ by 
comparing the measured value with the input one.   The simulation is run for 500 times. At EW=3 (typical value for 
$-0.5<log \oiii/\hb<1$ star-forming galaxy)  and $S/N$=3, 5, 7, 10, the errors in emission line width: $\sigma$ are 
0.19~dex,  0.08~dex, 0.06~dex, and 0.04~dex. For the worst case: EW=3, S/N=3, the error in $\sigma$ is 0.19~dex. 
For sources with higher \oiii/\hb\ and higher emission line width, the measurement are more reliable. 

\subsection{Intermediate-redshift data}
Our intermediate-redshift galaxy sample is based on observations from
the DEEP2 Galaxy Redshift Survey (hereafter DEEP2; Davis et al. 2003; Newman et al. 2013).
\footnote{http://deep.ps.uci.edu/}
The DEEP2 survey has a limiting magnitude of $R_{AB}$ = 24.1, and it
covers 3.2 $deg^2$ spanning 4 separate fields on the sky. The spectra span a
wavelength range of 6500-9100\AA\ at a spectral resolution of R$\sim$5000.
The DEEP2 DR4 include 52,989 galaxies.
For our study, we limit the redshift range to be 0.32 $<$ z $<$ 0.82,
to ensure the detection of \hb\ and \oiii\ in the DEEP2 wavelength coverage.
The sample size is cut to 12,739 galaxies.
The spectra are obtained with the DEIMOS spectrograph (Faber et al. 2003) at the Keck
Observatory and reduced with the pipeline\footnote{http://astro.berkeley.edu/$\sim$cooper/deep/spec2d/}
developed by the DEEP2 team at the University of California
Berkeley. 
All the DEEP2 footprints are observed by $Chandra$ Advanced CCD Imaging Spectrometer (ACIS-I)
with total exposures across all four XDEEP2 fields range from $\sim$10 ks to 1.1 Ms (Goulding et al. 2012;
Laird et al. 2009; Nandra et al. 2005).
The intermediate-redshift data is used for calibration of the new KEx diagnostic diagram at z$<$1.

\section{Classification using SDSS main galaxy sample}
\subsection{Classification}

\begin{figure*}
\begin{center}
\label{fig-1}
\includegraphics[width=18cm]{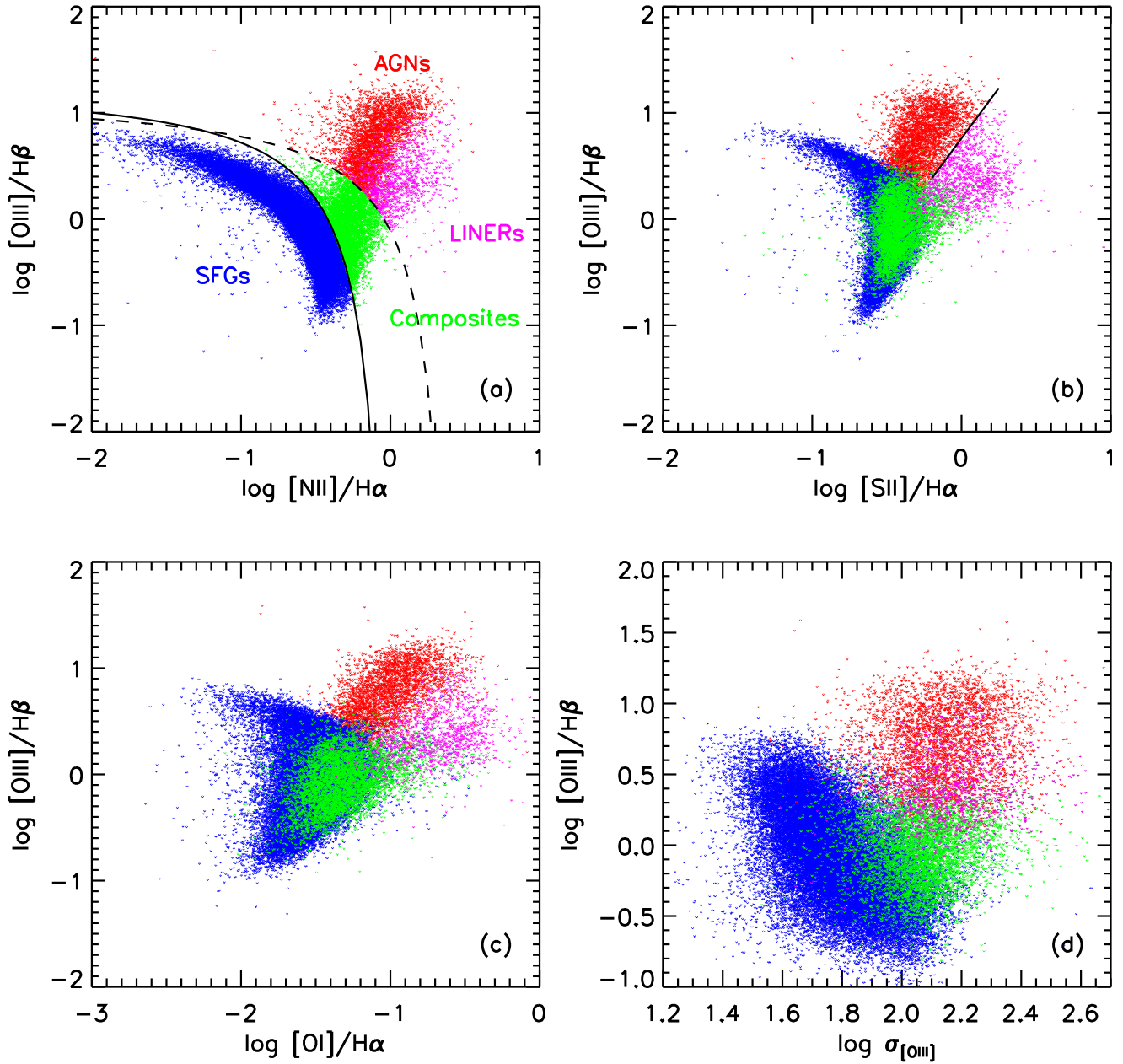} 
\caption{Pandel~(a)-(c): The BPT diagram for SDSS low redshift emission line galaxies classification.
The blue dots are star-forming galaxies, the green dots are composites, the red dots are Seyfert2s,
and the magenta dots are LINERs. The classification and color scheme are based on a combination of Panels (a) and (b). The solid and dashed 
lines in Panel~(a) are demarcation lines from Kauffmann et al. (2003) and Kewley et al. (2001). The solid line 
in Panel~(b) is from Kewley et al. (2006).
Panel~(d): The KEx diagram for all emission line galaxies from SDSS DR7 main galaxy sample. The legends are the same as Panel~(a). 
 KEx diagram is efficient in separating AGNs from SFGs. }
\end{center}
\end{figure*}

Since the BPT diagram needs at least \oiii, \hb, \nii\ and \ha\
for a classification, it is not applicable to sources at z$>$0.4
with optical spectrum alone. We propose a new diagnostic diagram: 
\oiii/\hb\ vs $\sigma_{\oiii}$ to diagnose the ionization source and physical properties of emission line galaxies. 
We call it the Kinematics-Excitation Diagram (KEx diagram) hereafter.
This approach shares similar logic to the work of Yan et al (2011) and Juneau et al (2011).

The dividing lines proposed by Kewley et al. (2001, 2006) and Kauffmann et al. (2003) are
used to classify the emission line galaxies into star-forming galaxies, composite galaxies,
Seyfert2s, and LINERS, as shown in Figure~1. We require the signal to noise ratio to be greater 
than 3 for \hb, \oiii, \oi, \nii, \ha\ and \sii\ lines
to ensure classification on the BPT diagram into sub-types. The \oi/\ha diagram is shown for reference and 
not taken into account in the classification. How the different types of galaxies
populate the KEx diagram is shown in Panel (d) of Figure~1. 

We plot different types of emission line galaxies on
\oiii/\hb\ vs $\sigma_{\oiii}$ plot in Figure~2 separately.
From left to right are star-forming galaxies, composite galaxies, LINERs and Seyfert2s.
In panel (a), the star-forming galaxies cluster around the lower-left corner on the diagram,
the boundary of the star-forming galaxies is clear and sharp.
We derive an empirical curve to follow the boundary:
\begin{equation}
\log \oiii/\hb =-2\times \log \sigma_{\oiii} +4.2
\end{equation}

This curve can be used to separate AGNs from star-forming galaxies.
The detailed distribution of BPT-classified galaxies on the KEx diagram
is given in Table~1. 
97\% (5674/5860) of the BPT-classified Seyfert2s (above Kewley01 line) and
35\% (5587/16003) of the BPT-classified composite sources lie above the new classification line and will be classified as AGNs by the KEx diagram.
98.8\% of BPT-classifed SFGs are classified as KEx-SFGs.
81\% of the KEx-classifed AGNs are BPT-classifed AGNs (above Kewley01 line),
90\% of the KEx-classifed SFGs are BPT-classifed SFGs. For all the
sources on the upper side of the line, 7.7\% are classified as star-forming galaxies in traditional BPT
diagram. 65.1\% (10,416/16,003) of BPT composites are in the KEx-SFG region. 46.9\% ((10,416+116+186)/(16003+998+5860)) 
of non-SF galaxies are on the KEx-SFG side. 
This means the new diagram is very efficient for selecting AGNs above Kewley01 line with high
completeness and low contamination rate.

One may notice that the composite galaxies cluster near the SFG-AGN dividing line and
clearly separate from Seyfert2s. We draw a horizontal line:
\begin{equation}
\log \oiii/\hb =0.3
\end{equation}
to cut out a region dominated by composites..
Above this line, the galaxies are mainly LINERs and Sy2s and we call it
KEx-AGN region. Below the log \oiii/\hb =0.3 line while 
above the SFG-AGN line, composite galaxies dominate, and we name it
KEx-composite region.
Precisely, there are 5,074 composites in KEx-composite region. In this region, there are only 335 Sy2s, 424 LINERs,
and 882 star-forming galaxies making up 24.4\% of non BPT composites in the KEx-composite region.

We can see in Figure~2 and Table~1 that 116/998 LINERs are classified
as star-forming galaxies on the KEx diagram,  426/998 are
KEx-composites,  and 458 are KEx-AGNs. LINER like emission could be produce by Low-luminosity AGNs (Ferland \&
Netzer 1983; Halpern \& Steiner 1983; Groves et al. 2004b; Ho 2008),
post-AGB stars (Binette et al. 1994; Yan \& Blanton 2012; Singh et
al. 2013), fast shocks (Dopita \& Sutherland 1995), photoionization by the hot X-ray-emitting
gas (Voit \& Donahue 1990; Donahue \& Voit 1991), or thermal
conduction from the hot gas (Sparks et al. 1989). Despite the ionization origins are diverse, the host
galaxies of LINERs are massive, making them only weakly
overlap with the star-forming galaxies on the KEx diagram. Since our
KEx diagram does not include the information of low-ionization lines,
the LINERs are not well separated from AGNs.

We note that only 1/3 of the BPT-classified composites galaxies are in
the KEx-composite region, while most of the remaining 2/3 are in the
KEx-SFGs region. Only a small fraction are in the KEx-AGN region. This may be becausethe
BPT-classified composite galaxies have a diverse origin too. They could
be relatively weak AGNs (Kauffmann et al. 2003; Yuan et al. 2010;
Ellison et al. 2011), shock heated (e.g., Rich et al. 2014. ), or
\hii\ region with a special physical condition (Kewley et
al. 2001). Trouille et al (2011) showed that the
composites are most similar to AGNs in their TBT diagram and
show not only photoionization properties like AGNs but also an
excess X-ray emission relative to the infrared emission,
indicating non stellar processes. This is a more likely scenario
than shocks or varying \hii\ region conditions. In the case of
shocks, even galaxies with a lot of regions locally dominated
by shocks have overall line ratios that place them in the
BPT-SFG region instead of the composite region (Rich et al 2011).
Some composites may have an intrinsically less
luminous AGN, and that the NLR gas is moving slower, both in rotation and in outflow.
This means our KEx diagram, which is successful in separating strong
AGN from SFGs, may not have enough diagnostic power to
pick out weak AGNs who have low contrast in both the \oiii/\hb\ ratio
and the kinematics relative to SFGs. KEx diagram also suffers from the mix of composites 
with other populations, especially the KEx-SFG, but to a lesser degree 
than some of the alternative diagrams.

In Figure~3, we plot KEx-AGN, KEx-composites, and KEx-SFGs on
the BPT diagram. The KEx-composites cluster around the composite region on the BPT diagram.
It is possible that the KEx diagram can be used to further
diagnose the real nature of the composite galaxies. 
Panel~(b) in Figure~2 also suggest that the \oiii\ line width could
potentially be used to further constrain the nature of
composite galaxies. For example, broad emission line width
is usually regarded as a tracer of shock (e.g., Rich et al. 2011,2014). 
The different line width between sub-population of composites could potentially be used 
to constrain the relative importance of AGN and SF processes. 
We leave the exploration of the power of the KEx diagram to diagnose sub-classes for future studies.

\begin{figure*}
\begin{center}
\label{fig-1}
\includegraphics[width=18cm]{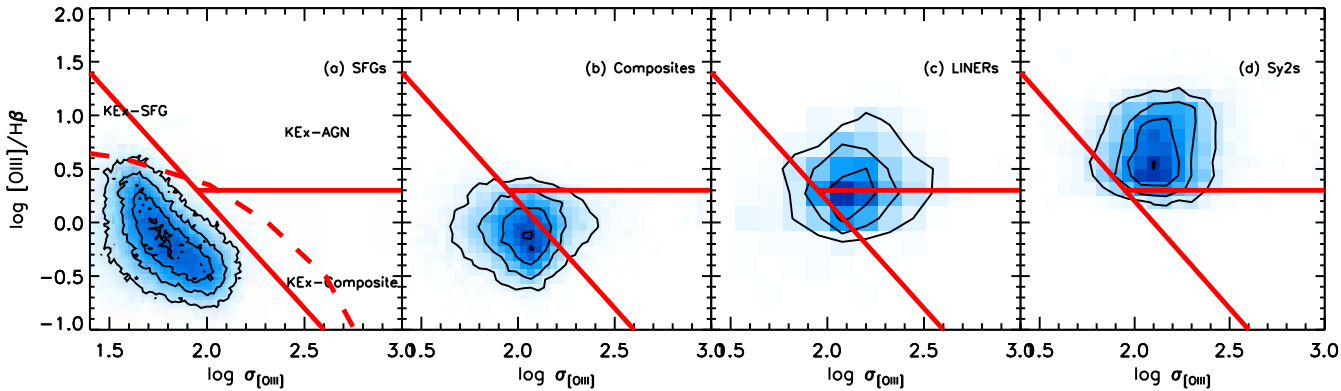} 
\caption{Panel~(a): The blue contours are the distribution of BPT-classified star-forming galaxies on the KEx diagram. The lowest level of the contour is 97 percentile
The solid lines are the dividing lines to classify emission
line galaxies into KEx-AGNs, KEx-composites, and KEx-SFGs. The dashed line is the Tully-Fisher Relation
prediction of the most luminous galaxies, as shown in Figure~5. Panel~(b)-(d): The distribution
of BPT-classified composites, LINERs and Seyfert2s on the KEx diagram. The solid lines are the same as Panel~(a).}
\end{center}
\end{figure*}

\begin{figure*}
\begin{center}
\label{fig-1}
\includegraphics[width=9cm]{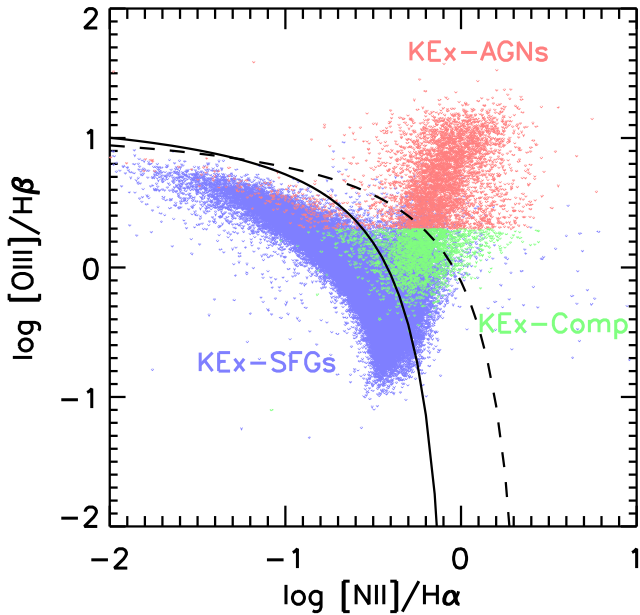} 
\caption{The distribution of KEx-AGN (pink), KEx-composites (green)
  and KEx-SFGs (purple) on the BPT diagram.  }
\end{center}
\end{figure*}

\subsection{Why the KEx diagram works?}

\begin{figure*}
\begin{center}
\includegraphics[width=16cm]{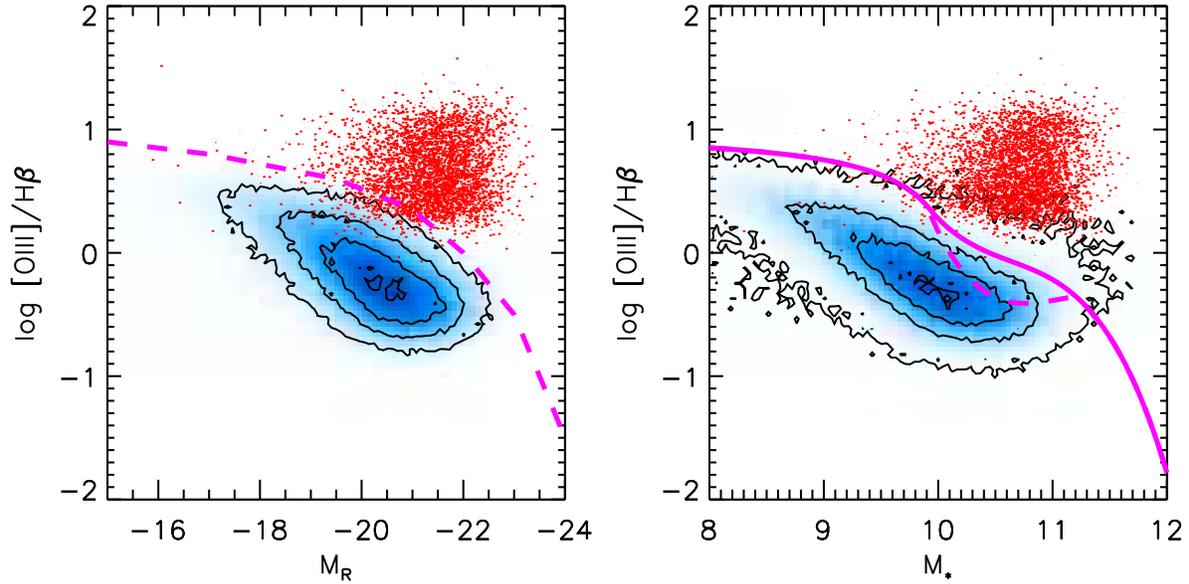} 
\caption{Left panel: \oiii/\hb\ vs. R band absolute magnitude. The blue contours are star-forming galaxies and the red dots are AGNs.
The lowest level of the contour is 90 percentile. The dash line is the boundary line we draw around the distribution.
Its corresponding line on the KEx diagram after Tully-Fisher Relation transformation is shown in left panel of Figure~2.
Right panel: \oiii/\hb\ vs. stellar mass. The blue contours are star-forming galaxies and the red dots are AGNs. The lowest levels is 90 percentile. The solid 
and dashed lines are the demarcation lines of MEx diagram (Juneau et al. 2011).  AGNs and SFGs are not separated as far on these 2 diagrams 
as on the KEx diagram.  }
\label{o3hb_trend.fig}
\end{center}
\end{figure*}

\begin{figure*}
\begin{center}
\includegraphics[width=16cm]{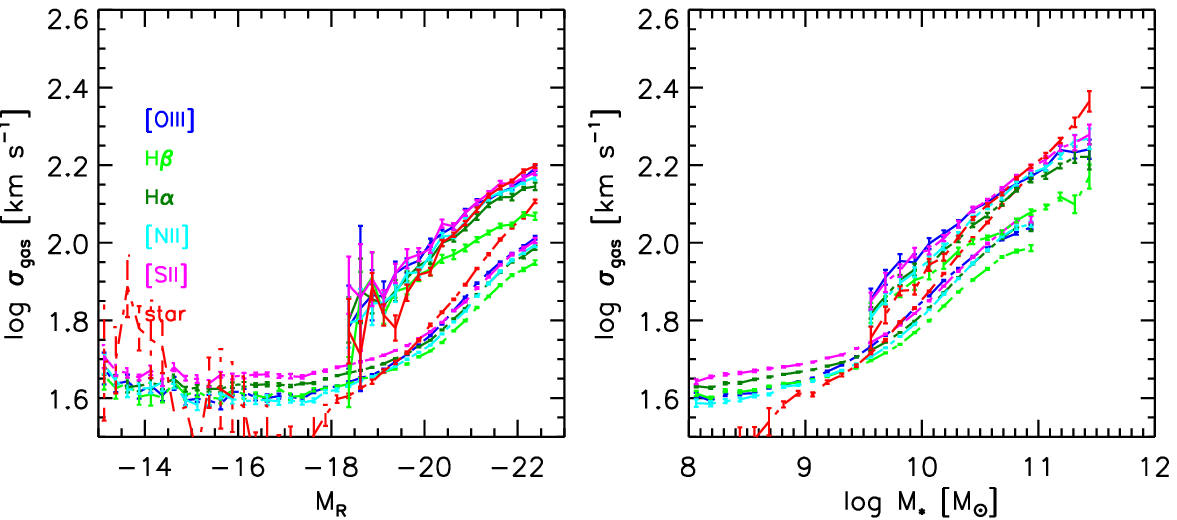} 
\caption{Left panel: Widths of different narrow lines against R band absolute magnitude. 
The solid lines and dash-dotted lines are median value of $\sigma$ at given R band absolute magnitude for Sy2s and star-forming galaxies respectively. 
Sy2s are shown as the solid line and star-forming galaxies
are represented by the dash-dotted line. Different color represent different emission lines and
the description of the legends are shown at the left-up corner.
Right panel: The median $\sigma$ of different narrow lines against stellar mass.
The legend is the same as the left panel. At fixed R band luminosity/ Stellar mass, AGNs show 0.15/0.12 dex higher 
$\sigma_{\oiii}$ than SFGs. This enhancement helps AGNs separate further from SFGs on the KEx diagram.  }
\label{sigma_trend.fig}
\end{center}
\end{figure*}

Comparing to the widely used BPT diagram, the KEx diagram uses $\sigma_{\oiii}$
instead of the \nii/\ha\  ratio as the horizontal axis to diagnose the ionizing source and physical properties of emission line galaxies.
We showed that it gives very consistent result with the BPT diagram.
Here we discuss the physical reasons behind the diagram and why it works.
$\sigma_{\oiii}$ in principle traces the motion of the gas.
To first order, this motion is determined by the gravity of the galaxy.
The emission line width correlate well with the stellar velocity dispersion in
almost all types of emission line galaxies (e.g, Nelson 2000; Wang \& Lu 2001;
 Bian et al. 2006; Chen et al. 2008; Komossa et al. 2007, 2008; Dumas et al. 2007; Greene \& Ho 2005; Ho 2009). 
In AGNs, several additional sources of line broadening may at work in addition to the stellar kinematics (Greene \& Ho 2005). 
So the basic principle behind the KEx diagram is the different kinematics of emitting gas in AGNs and star-forming galaxies.

The boundary between star-forming galaxies and AGNs on the KEx indicates there is a
maximum $\sigma_{\oiii}$ for star-forming galaxies.
In local star-forming galaxies, \oiii\ comes from the \hii\ region, which is more
correlated with the kinematics of the disk.  It was found that the width of narrow emission lines of star-forming galaxies could efficiently
trace the maximum rotation velocity of a galaxy (Rix et al. 1997; Mall\'{e}n-Ornelas et al. 1999; Weiner et al. 2006; Mocz et al. 2012), so
the line width reflects how fast the disk is rotating.
The rotation speed is directly linked to the luminosity of the galaxy through
the Tully-Fisher relation (TFR, Tully \& Fisher 1977). The most luminous spiral
galaxies also rotate the fastest. A reasonable hypothesis is that the galaxies near
the maximum $\sigma_{\oiii}$ boundary in the KEx diagram are the most
luminous ones. We plot all the star-forming galaxies in our sample on
the \oiii/\hb\ vs. R band absolute magnitude in Figure~4.
The star-forming galaxies cluster in the left-down corner of the diagram, and show a clear boundary.
We draw a magenta boundary curve to define the maximum luminosity a
star-forming galaxy can reach at a given \oiii/\hb. Using
the Tully-Fisher relation obtained by Mocz et al. (2012), we convert the luminosity in the curve
to $\sigma_{gas}$, and obtain a \oiii/\hb\ vs. $\sigma_{gas}$ curve, which is shown as the dashed line in Figure~2 (a).
We can see that the dashed line covers the regions where SFGs and Composites locate,  separating Seyfert2s from SFGs and Composites. 
This illustrates that SFGs and Composites are consistent with the TFR relation prediction, while AGNs are located outside the locus.

We overplot the distribution of Seyfert2s (red dots) on the
\oiii/\hb\ vs. $M_r$ (absolute petrosian magnitude) and
\oiii/\hb\ vs. $M_*$ (stellar mass) diagrams (MEx Diagram, Juneau et al. 2011) in Figure~4. The magenta lines in the right panel are the MEx dividing lines.  
The sources above the upper curve are MEx-AGN and the region between the solid upper curve and dashed line is the MEx-composite region. The contours are 5, 40, 70, 95 percentiles. 
The stellar mass is drawn from the MPA-JHU catalog. Seyfert2s occupy the bright and massive end of Figure~4.
This is consistent with previous findings that Seyfert2s reside exclusively in massive,
luminous galaxies (e.g., Kauffmann et al. 2003).
Even though the host of Seyfert2s are luminous and massive, the overlap between Seyfert2s and
star-forming galaxies on these two plots is large compared with that on the KEx diagram. 
4.2\% (246/5860) of Sy2s are on the MEx-SF side, while 3.2\% (186/5860) of Sy2s are on the KEx-SF side. 
On the \oiii/\hb\ vs $M_R$ plot,   7.7\% (450/5960) of Sy2s lie on the SFG side of the dividing curve shown in the left panel of Figure~4. 
The fraction of BPT-classified Seyferts that are mis-classified as SFGs are higher on these two diagrams than on the KEx diagram. 

There is an additional enhancement in $\sigma_{\oiii}$ at fixed luminosity or stellar mass for AGNs relative to SFGs. This enhancement helps AGNs separate further from SFGs on the KEx diagram. 
In Figure~5, we plot the median $\sigma_{gas}$ derived from five emission lines against $M_r$ and stellar mass in the left and right panels. We focus on the $\sigma_{\oiii}$ (blue line) first and discuss other emission lines later. The errors are calculated using the bootstrap method.
We can see that at a given luminosity where both AGNs and SFGs cover,
the $\sigma_{\oiii}$ of Seyfert2s is on average 0.15~dex higher than star-forming galaxies.
And at a given stellar mass, Seyfert2s are 0.12~dex higher than star-forming galaxies.
These offsets are critical to the clean separation between Seyfert2s and star-forming galaxies on the KEx diagram. Other emission lines gave similar results.

One reason of the difference in $\sigma_{\oiii}$ may be that the
emission lines in Seyfert2s are produced by gas in the narrow line
region which extend into the bulge and the emission line in star-forming
galaxies are emitted by gas in the \hii\ region in the disk.
The matter distribution and kinematics are very different in the disk and bulge of a galaxy
It is shown in Catinella et al. (2012) that at a given luminosity or baryon mass,
the disk dominated galaxy show 0.1~dex smaller $\sigma_*$ (measured from the SDSS 3" fiber)
than the bulge dominated galaxies. At a given mass, the bulge dominated galaxies, which are more
concentrated, show a higher $\sigma_*$ than the disk dominated galaxies.
Besides the difference in host galaxies, several other physical reasons may be
involved in the fact that Seyfert2s have broader emission lines.
Greene \& Ho (2005) found that the excess \oiii\ line width relative to the stellar or lower ionization lines kinematics is about 30-40\% (0.11-0.15 dex), with small variations depending on AGN luminosity, AGN Eddington ratio, SFR, etc. When considering only the core of \oiii, the excess in \oiii\ line width goes away (Green \& Ho 2005, Komossa et al. 2008). This supports that the \oiii\ core is produced by ionized gas in bulge while another source of broadening related to AGN (such as possibly winds or outflows from accretion disk) is at work. 
Radio jet may play a part in broadening the emission line too (Mullaney et al. 2013). 

The differences in $\sigma_{gas}$ and $\sigma_{*}$ between AGNs and SFGs are more pronounced against $M_r$ than against M*. In particular, $\sigma_*$ is significantly higher in AGN hosts at a fixed $M_r$ but not so different at a fixed stellar mass. This means that AGN hosts have higher mass-to-light ratios, which can be interpreted as having more important bulge components. More massive bulges host more massive black holes, and therefore be detectable as Seyfert 2s down to lower Eddington ratios. Conversely, lower mass AGNs are only identified as Seyfert 2s for comparatively higher Eddington ratios and their \oiii\ line width excess could be more pronounced relative to other lines than for higher mass AGNs if the Eddington ratio is the driving factor for additional broadening (e.g., Greene \& Ho 2005; Ho 2008).

In summary, the boundary between star-forming galaxies/composites and AGNs on the KEx diagram is defined by the Tully-Fisher relation of the most luminous and massive galaxies. The AGNs reside in luminous and massive galaxies and at a given luminosity/stellar mass, their $\sigma_{\oiii}$ are 0.15/0.12~dex 
higher than the star-forming galaxies. These effects make the KEx diagram an efficient classification
tool for emission line galaxies.

\section{KEx Diagram Calibration at 0.3$<$z$<$1 }
The main purpose of the KEx diagram is for emission line galaxies classification
at high redshift. We have demonstrated the KEx diagram could successfully separate emission line galaxies
in local universe, mainly due to AGN occur in massive galaxies with high bugle-to-disk ratio and AGN have extra broadening due to outflow . At high redshift,  the properties of galaxy and AGN host are different.  $\sigma_\oiii$ could be higher because galaxies were more gas rich, there were more unstable disks with high "$\sigma$/V" (Papovich et al. 2005; Reddy et al. 2006; Tacconi et al. 2010; Shim et al. 2011) and AGN were more luminous so had higher Eddington ratios.  Besides, the \oiii/\hb\ of the star-forming galaxies would be higher too. Galaxies at $z\sim1.5$ have typically higher \oiii/\hb\
ratios than z$<$0.3 galaxies (e.g., Liu et al. 2008; Brinchmann et al. 2008;
Trump et al. 2011, 2012; Kewley et al. 2013a,b). The physical properties in \hii\ region and AGN NLR could be different from the local
universe as well (Kewley et al. 2013a,b).  There is also evidence for AGN in relatively low mass hosts at higher redshifts (Trump et al 2011). 

To calibrate, we need a sample of AGNs and star-forming galaxies that are already classified .  
There are several methods for calibration:
Firstly, X-ray identification may act as an independent reference for classification calibration
(Yan et al. 2011; Juneau et al. 2011, 2013) even though with some drawbacks.
The X-ray AGNs and optical AGNs may not be the same population (Hickox et al. 2009) and the
sources with reliable deep X-ray data are limited. Besides, X-rays surveys are
less sensitive to moderate-luminosity AGNs in galaxies
of lower stellar masses (Aird et al. 2012) or heavily obscured Compton-thick
AGNs. Yan et al. (2011) estimated at \lbol$> 10^{44}erg s^{-1}$,
about 2/3 of the emission-line AGNs with 0.3$<$z$<$0.8 and $I_{AB}<22$ will
not be detected in the 2-7~keV band in the 200 ks
Chandra images due to absorption and/or scattering of the X-rays in the EGS field.
We use DEEP2 data and X-ray identification for KEx calibration in Section~4.2. 
DEEP2 contain both star-forming galaxies and AGNs, mostly star-forming galaxies. This 
can help us constrain our calibration on the SF side. 
Secondly, If NIR spectra are available, the \nii\ and \ha\ emission lines could be used for optical classification using the BPT diagram as a corner stone.
The use of the BPT at higher redshift remains potentially valid, but that it needs to be further verified and the sample size is limited as well(Trump et al. 2012, Kewley et al. 2013b; Juneau et al. 2014). At z$\sim$1, Type2 AGN sample is very limited, because 
Type2 AGNs can only be identified using BPT diagram to z$\sim$0.4 using only optical 
spectrum, and the NIR spectroscopy sample which enable the BPT diagram to extend to z$\sim$1 is limited. 
Thirdly, there are many Type1 AGNs from SDSS in the 0.3$<$z$<$1 range, because the broad lines can be identified using bluer wavelength 
range, and the volume covered by SDSS is large. Type 1 AGNs can be identified by their blue color or/and broad emission lines. We use them as independently identified AGNs to perform a sanity check of the KEx classification when considering only the narrow line components in Section~4.1. We make the assumption that the Type1 and Type2 sources have identical narrow line features in the frame of unification model (Antonucci et al. 1993). Some differences in narrow lines indeed exist due to NLR stratification, outflow, or
narrow line Baldwin effect (Veilleux et al. 1991; Zhang et al. 2008, 2011; Stern \& Laor 2013),
but the assumption is valid grossly.  

\subsection{Calibration using Type1 AGNs}
Our KEx diagram uses only \hb\ narrow component and \oiii\ emission lines so it is straightforward to use type1 AGNs to calibrate our KEx diagram.
We first use a sample of low-z(z$<$0.3) type1 AGNs from SDSS DR7 main galaxy
sample for testing purpose. These sources have \ha\ broad component with significance greater than 3$\sigma$ as described in Section~2.1. 
We plot these sources in the KEx diagram in
the Panel (a) of Figure~6. The \oiii/\hb\ only include the narrow
component of \hb. These sources (4624) reside mostly in the KEx-AGN
and KEx-composite regions,
while only 414(9\%) of them are in the KEx-SFG region. 

We further plot a sample of intermediate-z Type1 AGNs selected from SDSS DR4 QSO catalog with 0$<$z$<$0.8 on the KEx diagram.
The sources in this sample have small contamination from host galaxies
in their optical light, and their properties and data reduction are described in
Dong et al. (2011). The detailed analysis of narrow line properties, especially the \oiii\
line, could be found in Zhang et al. (2011, 2013b). The bolometric luminosity range of these
sources is $10^{44} \sim 10^{47}$ $erg s^{-1}$. In left panel of Figure~6, we plot this sample
on KEx diagram in orange.  Almost all(96\%) of the sources lie in the
KEx-AGN region, and a small fraction of the Type1 sources(4\%) lie in the KEx-composite region.
Only 13(0.3\%) Type1 sources are classified as KEx-SFGs.
According to unification model, if these Type1 sources
are viewed edge-on, almost all of them would be rightly classified as AGNs. 
There are few points at z>0.3 and that they may be consistent with either no shift or a small shift of 0.1~dex. 

One may notice that some Type1 AGNs lie outside the low-z locus. Some sources have
higher \oiii/\hb\ line ratio and some have larger $\sigma_{\oiii}$.
These could be partly understood
by the orientation effect. The Type1 AGNs are found to have higher
ionization state than Type2 AGNs (Veilleux et al. 1991c; Schmitt et al. 2003a,b), and this is
because that high-ionization lines arise from regions closer to the nuclei
thus more likely to be blocked when viewed edge-on. The inclination effect may play a role in the higher width of \oiii\ emission here. \oiii\ emission line is
known to show blue-wing asymmetric profile (Heckman et al. 1981; Zhang et al. 2011)
and this is believed to be due to narrow line region outflows.
When the outflows are viewed in a face-on orientation,
we would see larger overall outflow velocity and this would lead to larger line width
at the same time.

\subsection{Calibration using the DEEP2 Survey}

\begin{figure*}
\begin{center}
\includegraphics[width=16cm]{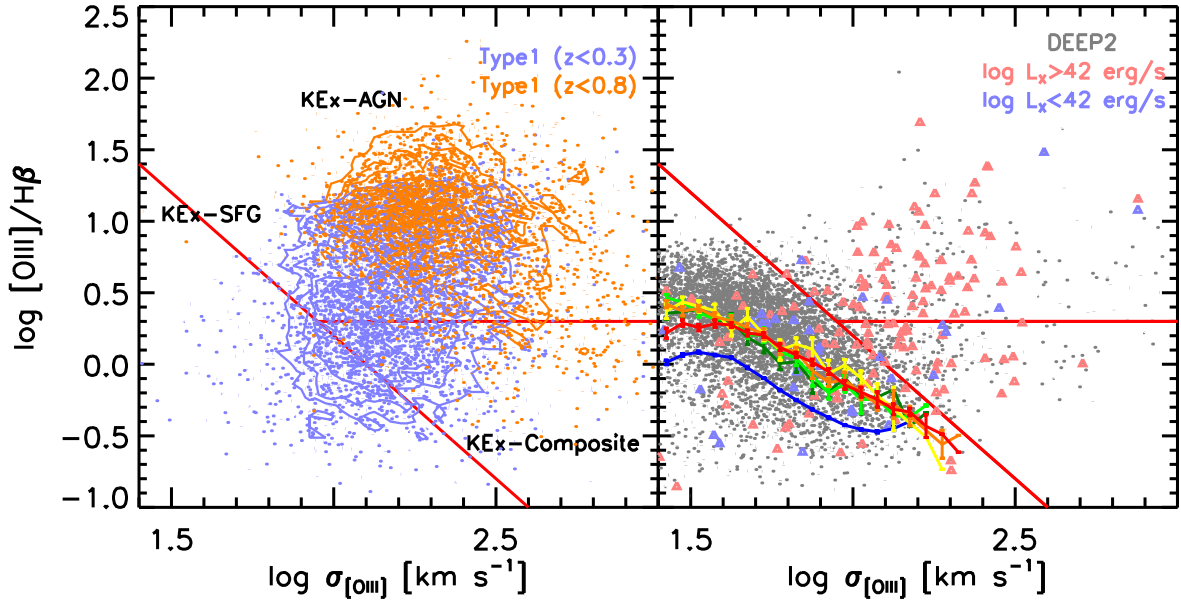} 
\caption{Left panel: 
Left panel: The purple dots and contours are sources with broad \ha\ emission line from SDSS DR7 main galaxy sample.
The orange dots and contours are the Type1 AGNs from Dong et al. (2011).
The solid lines are KEx demarcation lines. Type1 AGNs are located mainly in the KEx-AGN and KEx-Composite regions. 
Right panel: The gray dots are intermediate redshift galaxies from DEEP2 survey. The
pink dots are X-ray AGNs, selected using $log$ $L_X(2-10keV)> 42$ $erg/s$. The purple dots are galaxies with $log$ $L_X(2-10keV)< 42$ $erg/s$,
X-ray star-forming galaxies according to our criteria. X-ray AGNs are located mainly in the KEx-AGN and KEx-Composite regions, 
while X-ray SFGs are located in the KEx-SFG region. At z$<$1, the KEx 
diagram does not need re-calibration. The  blue, dark green, green, yellow, orange, and red 
lines are median \oiii/\hb\ at given $\sigma_{\oiii}$ for $z\sim0.1$
(from SDSS), $0.3<z<0.4$, $0.4<z<0.45$, $0.5<z<0.6$, $0.6<z<0.7$, $0.7<z<0.8$ (from DEEP2)  KEx-SFGs respectively. The seemingly offset between 
z$\sim$0.1 and higher redshift is consistent with aperture effect. }
\label{hiz.fig}
\end{center}
\end{figure*}

Our intermediate-redshift galaxy sample is based on observations from
the DEEP2 Galaxy Redshift Survey. Most of the galaxies in DEEP2
surveys are star-forming galaxies, as indicated in X-ray study (Goulding et al. 2012;
Laird et al. 2009; Nandra et al. 2005), and there are also many X-ray
AGNs in this sample.
Even though using the X-ray data to calibrate the KEx suffer some problems as
discussed in previous subsection and in other works (Hickox et al. 2009;
Yan et al. 2011; Juneau et al. 2011, 2013), it is interesting to check
if the X-ray and the KEx classification are consistent and what causes the
differences  An X-ray luminosity threshold: $L_{2-10keV} > 10^{42} erg s^{-1}$ is adopted. There is no star-forming galaxies with X-ray luminosity higher than this value in local universe.The sensitivity of the X-ray data will not result in
mis-classification of AGNs and star-forming galaxies,
even though faint sources are missed in shallow areas. However, weak AGNs with $L_{2-10keV} < 10^{42} erg s^{-1}$ exist even though they are more ambiguous to differentiate from star-forming or starbursting galaxies with X-ray emission without additional information.
For DEEP2 X-ray data, the sensitivity of the shallowest data could ensure the
detection of luminous X-ray sources ($L_{2-10keV} > 10^{42} erg s^{-1}$).

The \oiii/\hb\ vs $\sigma_{\oiii}$ for DEEP2 are plotted in green triangles in right panel of Figure~6.
We use the same method described in Section~3 for spectral fitting.
We apply a S$/$N cut of 3 to \hb\ and \oiiil. 7,866 sources satisfy this criteria.
We convert the hard 2-8~keV X-ray flux to rest-frame 2-10keV luminosities ($L_X(2-10 keV)$)
by assuming a power-law spectrum with photon index ($\gamma=1.8$).
The sources with $L_X(2-10 keV)>10^{42} erg s^{-1}$ are classified as X-ray AGNs, and
sources with $L_X(2-10 keV)<10^{42} erg s^{-1}$ are classified as star-forming galaxies.
We caution $L_X(2-10 keV)>10^{42} erg s^{-1}$ sources may be star-forming galaxies at high-z, due to higher SF activity in the early universe. 
Many $L_X(2-10 keV)<10^{42} erg s^{-1}$ may be AGNs but dim intrinsically or due to obscuration.
The X-ray sources are plotted in pink and purple triangles in right panel of Figure~6. We can see that most of the X-ray
AGNs/star-forming galaxies are consistently classified as optical AGNs/star-forming galaxies.
Out of the 93 X-ray AGNs, 48 (52\%) are classified as KEx-AGNs,
18(19\%) are KEx-composite, and 26 (29\%) are KEx-SFGs.

Out of the 83 X-ray starbursts in our sample, 19 are KEx-AGNs,
14 are KEx-composite, and 49 (59\%) are KEx-SFGs.
In Juneau et al. (2011), for the X-ray starbursts, 50\% (8/16) are classified
as MEx-SFGs, while 19\% (3/16) are in the intermediate region and the remaining
31\% (5/16) reside in the AGN region. The two results are consistent.

On the other hand, we notice that some sources are optically classified as star-forming galaxies 
but have very powerful X-ray emission, indicating
harboring an active nuclei.
29\% of X-ray AGNs are classified as star-forming galaxies on the KEx diagram.
In Juneau et al. (2011), 20\% of their X-ray AGNs are classified as MEx-intermediate, and 15\% are MEx-SFGs.
Considering the intermediate region is mixed with
star-forming region on MEx diagram, our result is consistent with theirs.
Yan et al. (2011) found 25\% of their X-ray AGNs reside in star-forming region of their
optical classification diagram which replaces the \nii/\ha\ in the BPT diagram with rest-frame U-B color.
This is consistent with our result too. Castello-Mor et al. (2012) studied the sources with $L_X(2-10 keV)>10^{42} erg s^{-1}$ but
classified as star-forming galaxies on the BPT diagram. These sources have large thickness parameter ($T=F_{X}/F_{\oiii}$),
large X-ray to optical flux ratio ($X/O>0.1$), broad \hb\ line width, steep X-ray spectra, and
display soft excess. These mis-matches illustrate neither X-ray or optical classification are complete.  Different classification schemes are complementary to each other. 
At z$<$1, the evolution in $\sigma_{\oiii}$  and \oiii/\hb\ is not large so we don't need to shift the dividing line.

\section{Calibration at z$\sim$2}

\begin{figure*}
\begin{center}
\includegraphics[width=8cm]{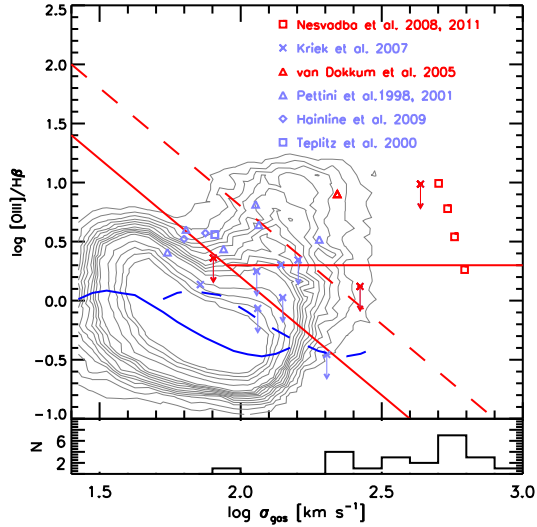} 
\caption{ z$\sim$2 galaxies from literature in the KEx diagram. The references where the
data come from is shown at the right-up corner. The purple color denotes the sources are star-forming galaxies, while
the red color denotes they are AGNs. Upper limits are shown with downward arrows.
The red lines are the same demarcation lines as Figure~2. The dashed line is the KEx-SFGs dividing
line shifted 0.3~dex right-ward. The solid purple line is the
\oiii/\hb\ vs $\sigma_{\oiii}$ relationship for $z\sim0.1$ KEx-SFGs
from DEEP2, and the dashed purple line is derived by shifting the relationship for $z\sim0.1$ KEx-SFGs
0.3~dex right ward. The \oiii\ width distribution of z$\sim$2 QSOs from Netzer et al. (2004) is shown in the lower
panel. The demarcation lines needs to be shifted due to the evolution of kinematics and \oiii/\hb\ at z$\sim$2.  }
\label{z2.fig}
\end{center}
\end{figure*}

In this section, we discuss extrapolating the KEx diagram to redshift greater than 2, and
use z$>$2 emission line galaxies to test the validation of the diagram.
The $2<z<4$ epoch is critical to galaxy formation and evolution. During this epoch, the Hubble sequence
was not fully established (Kriek et al. 2009; F\"{o}rster Schreiber et al. 2006, 2009),
but the bimodality was in place (Kriek et al. 2009). It is at this redshift that
the star-formation density and AGN activity peak (e.g., Barger et al. 2001;
Elbaz \& Cesarsky 2003; Di Matteo et al. 2005; Hopkins 2004; Hopkins et al. 2006; Hopkins \& Beacom 2006).
There are thousands of emission line galaxies discovered
at this redshift range, and most of them are detected using the dropouts techniques
(e.g., Steidel et al. 1996, 1998, 2003, 2004; Pettini et al. 1998, 2001),
or color-selection (Franx et al. 2003; Daddi et al. 2004; Kong et al. 2006 ).

At z$<$1, the KEx can be used to successfully separate 
AGNs and star-forming galaxies mainly due to 3 reasons:
First, AGNs reside in the most luminous and most massive galaxies.
Second, the Tully-Fisher relation which define the boundary between star-forming galaxies/composites and AGNs is valid.
Third, the AGNs have systematic higher $\sigma_{gas}$ than star-forming galaxies at fixed luminosity
and stellar mass.
The first condition is likely to be true for AGNs up to z$\sim$3 (Xue et al. 2010; Mullaney et al. 2012) because 
the moderately luminous AGN fraction depends strongly on stellar mass but only weakly on redshift.
The TFR, however, is likely to evolve with redshift.

Unlike local disk galaxies who is totally rotation-supported with $V_{rot}/\sigma=10\sim 20$(e.g. Dib, Bell \& Burkert 2006),
a large fraction of high redshift star-forming galaxies have velocity dispersion comparable or even larger than rotation velocity(F\"{o}rster Schreiber et al. 2006, 2009; Wright et al. 2007; Law et al. 2007; Genzel et al. 2008; Cresci et al. 2009; Vergani et al. 2012). Since we use the integrated emission line profile, the increase in velocity dispersion further broaden the emission line.

Even after we have extracted the rotation velocity
using the Integrated Field Spectroscopy (IFS) for these high-redshift galaxies, the derived Tully-Fisher relation is still different from the local
well-defined relation. The rotation speed of z$\sim$2 galaxies is $\sim$0.2~dex higher than galaxies of similar
stellar mass in local galaxies (Cresci et al. 2009; Gnerucci et al. 2011). Meanwhile, the \oiii/\hb\ of star-forming galaxies is known
to be higher at high redshift (Shapley et al. 2005; Erb 2006a; Groves et al. 2006; Liu et al. 2008;
Brinchmann et al. 2008; Shirazi et al. 2014). Thus, we expect that the KEx diagram, particularly the separation boundary of AGN and star-forming galaxies,
evolve to higher $\sigma_{gas}$ or higher  \oiii/\hb\  with redshift because the TFR and  \oiii/\hb\ redshift evolution,
and galaxies at high redshift have larger velocity dispersions, 

To explore these effects in detail and test the application of the KEx diagram at z$\sim$2, we compile a sample of
Lyman-break galaxies (LBGs) and color-selected BzK galaxies at $z>2$ who have \oiii\ and \hb\
emission line ratio and gas velocity dispersion measurements from literature.
Our sample include 10 galaxies (3 are AGNs) from Kriek et al. (2007), 1 LBG from Teplitz et al. (2000),
6 LBGs from Pettini et al. (1998, 2001), and 2 gravitationally-lensed star-forming galaxies from Hainline et al. (2009).
The AGN fraction of this sample is very limited because LBGs have very low AGN fraction (about 3-5\%, Erb et al., 2006b, Reddy et al., 2005, Steidel et al., 2002).

Figure~7 shows the $z>2$ LBGs are not confined to the KEx-SFGs region of local galaxies. Most of them lie in the KEx-AGN or KEx-composite region. 
If we shift the dividing line 0.3~dex to the right, most of the galaxies are rightly classified. 
Out of the 0.3 shift, the evolution of TFR could contribute 0.2~dex (Cresci et al. 2009; Gnerucci et al. 2011; Vergani et al. 2012; Buitrago et al. 2013). 
Besides, the high redshift star-forming galaxies have higher velocity dispersion, which is not included in the evolution of TFR. 

To test if AGNs still show different kinematic from star-forming galaxies at this redshift, we compile radio galaxies
that are confirmed to be AGNs and see how they distribute in the KEx diagram.
We use 4 radio AGNs from Nesvadba et al. (2008, 2011) and CDFS-695: A shock or/and AGN from
van Dokkum et al. (2005). They clearly separate from the star-forming galaxies on the KEx diagram after shifting the boundary by 0.3~dex.
We further check the \oiii\ width distribution of a sample of z$\sim$2 QSOs from
Netzer et al. (2004) and plot the histogram on lower panel of Figure~7. Only one source
has \oiii\ width less than 100km/s, and most of the QSOs have $\sigma_{\oiii}\sim300km/s$.
There is no doubt that these QSOs would have high \oiii/\hb\ ratio even though we do not have their detailed values. Therefore, they are
likely to separate from the z$\sim$2 star-forming galaxies.
Strong outflow driven by AGN is reported in the Radio Galaxies (Nesvadba et al. 2008, 2011),
so the large line width in \oiii\ is at least partly due to the outflow as discussed in Section~3.2.
Judging from our empirical results, the KEx diagram is likely to
work after shifting the dividing line 0.3~dex to the right at z$\sim$2. More data is needed to
better constrain the boundary.


\section{Other Relevant Issues}

\subsection{Comparison with previous classification diagrams}
Tresse et al. (1996) and Rola et al. (1997)
proposed to use EW(\oii), EW(\oiii) and EW(\hb) for galaxy classification at high redshift.
Stasi\'{n}ska et al. (2006) studied using \oii\ for galaxy classification, and even proposed a method that uses $D_n(4000)$, EW(\oii), and
EW(\neiii) (DEW diagram) to select pure AGNs with z$<$1.3 using only optical spectrum.
But different types of galaxies overlap with each other severely on these classification diagrams.

Trouille et al. (2011) proposed to use $g-z$, \neiii, and \oii\ to clearly separate AGNs
from star-forming galaxies. This method is very efficient in separating different types of
galaxies, but the \neiii\ emission line is weak even in AGNs. Thus this diagram requires
high signal-to-noise ratio spectra of galaxies for reliable classification. This is particular hard for
high redshift objects which are usually faint.

Our KEx diagram has a similar logic as the Color-Excitation (CEx) and Mass-Excitation (MEx) diagrams
proposed by Yan et al. (2011) and Juneau et al. (2011). The CEx diagram makes use of the fact
that AGNs reside in red or green galaxies in local galaxies. But this is likely
to be wrong at higher redshift (Trump et al. 2012). The MEx diagram and the one using rest-frame H-band
magnitude (Weiner et al. 2006) is based on the fact that AGNs are harbored by massive galaxies.
This is more robust at high redshift due to the AGN downsizing effect.
Trump et al. (2012) tested the validation of these two diagrams at z$\sim$1.5 and
found that the MEx remains effective at z$>$1 but CEx needs a new calibration.
As discussed in Section~3.2, we can separate AGNs and star-forming
galaxies because AGN reside in massive galaxies, and have $\sigma_{\oiii}$ 0.12~dex higher than star-forming galaxies of similar
stellar mass as an enhancement. This enhancement makes the KEx diagram more
efficient at separating AGNs from star-forming galaxies.  
One advantage of the KEx is the requirements for only a spectrum that covers a small spectral range to obtain all required quantities. 
The KEx diagram only requires \oiii\ and \hb\ lines for a robust classification. 

\subsection{Comparison of different narrow line widths}
In KEx diagram, we use \oiii\ emission line width for diagnostic. In order to check if different lines have different width,  in Figure~5 we plot the median value of $\sigma$ against R band absolute magnitude and stellar mass for different narrow lines using galaxies from SDSS DR7. We plot the stellar velocity dispersion ($\sigma_*$) in red line for reference. The $\sigma_{*}$ is stored in the SDSS spectrum file header. 32 K and G giant stars in M67 are used as stellar templates. These stellar templates are convolved with the velocity dispersion to fit the rest-frame wavelength range 4000-7000\AA\ by minimizing $\chi^2$. The final estimation is the mean value of the estimates given by the "Fourier-fitting" and "Direct-fitting" methods. 
We found that the SFGs show similar sigma for all emission lines. The Seyfert2s, however, show some systematics in line width. The \oiii\ line is broader
than other low ionization lines and recombination lines. This is expected, because
the \oiii\ emitting region is more concentrated due to its high ionization potential (Veilleux et al. 1991; Trump et al. 2012). 
Unexpectedly, \hb\ show the smallest line width, and the discrepancy is largest at the high luminosity high stellar mass end. 
One possibility for this discrepancy is the Balmer absorption fitting is not perfect. The incorrect absorption fitting affects 
resulting emission line flux and profile. Groves et al. (2012) found significant discrepancy in \hb\ when using CB07 for SDSS DR7 
and BC03 for SDSS DR4. The \ha, which should arises from the same emitting region of \hb, does not follow the behavior
of \hb\ but show the same trend as low-ionization lines because the absorption correction is much milder. 

When comparing the line width of emission lines and $\sigma_*$, we found that the
difference between $\sigma_*$  in star-forming galaxies and Seyfert2s of similar stellar mass is very small. 
At high stellar mass end, galaxies tend to have bulges, and those would contribute to increase the measured $\sigma_*$. 
However, if the ionized gas in SFGs still comes from the disk, then this component does not get the additional dispersion contribution from the bulge. 
Also, one can expect  $\sigma_{gas}<\sigma_*$ because the gas is more dissipative than the stars and can slow down dynamically (Ho 2009). 
We leave this topic for future studies.


\subsection{\oiii/\hb\ evolution}
It is found that the \oiii/\hb\ in star-forming galaxies gets higher at z$\sim$1-2 
(Shapley et al. 2005; Erb 2006a; Liu et al. 2008; Brinchmann et al. 2008; Hainline et al. 2009;
Wright et al. 2010; Trump et al. 2011; Shirazi et al. 2014). The reason
for this trend is under debate. Brinchmann et al. (2008)
found that the location of star-forming galaxies in the \oiii/\hb\ versus \nii/\ha\ diagnostic diagram
highly depends on their excess specific star formation
rate relative to galaxies of similar mass. They infer that an elevated ionization parameter U is
responsible for this effect, and propose that this is also the cause of higher \oiii/\hb\ in
high-redshift star-forming galaxies in the BPT diagram (Brinchmann et al. 2008; Shirazi et al. 2014) . 
Liu et al. (2008) argue that the high \oiii/\hb\ sources have higher electron densities and temperatures. 
It is also possible that AGN or shock increase the \oiii/\hb\ ratio (Groves et al. 2006; Wright et al. 2010). It is interesting to check how
\oiii/\hb\ evolves with redshift at given $\sigma_{\oiii}$ in the KEx diagram.

In the right panel of Figure~6, we plot the median \oiii/\hb\ at given $\sigma_{\oiii}$ for the SDSS z$<$0.33
galaxies and the DEEP2 galaxies on the KEx diagram. The red, orange, yellow, green, blue and purple
lines are median \oiii/\hb\ at given $\sigma_{\oiii}$ for $z\sim0.1$, $0.3<z<0.4$,
$0.4<z<0.45$, $0.5<z<0.6$, $0.6<z<0.7$, $0.7<z<0.8$ KEx-SFGs respectively.  We can see in Figure~6 that
the \oiii/\hb\ vs. $\sigma_{\oiii}$ relation does not evolve from z$\sim$0.3 to z$\sim$0.8, but these galaxies have on average
0.2~dex higher \oiii/\hb\ than the local galaxies at given $\sigma_{\oiii}$. However, we note that SDSS has a fixed aperture of 3" which acquires 
the light from the center of the galaxy , while DEEP2 spectra are obtained through long-slits, which enable them to include light from the outskirt of the galaxy. 
The outskirt of the galaxy have lower metallicity and larger rotation speed than the center. 
This would shift the  \oiii/\hb\ vs. $\sigma_{\oiii}$ relation in the right-up (higher \oiii/\hb, higher $\sigma_{\oiii}$) direction as we see . 
So our result are consistent with no evolution in the  \oiii/\hb\ vs $\sigma_{\oiii}$ relation from z=0-0.8.



\section{Summary and Conclusion}
We propose a new diagram, the Kinematic-Excitation diagram (KEx diagram), using the \oiii/\hb\ line ratio and the \oiiil\
emission line width ($\sigma_{\oiii}$) to diagnose the emissions of the
AGNs and the star-forming galaxies.
The KEx diagram uses only the \oiiil\ and \hb\
emission lines, thus it is a suitable tool to classify
emission-line galaxies (ELGs) at higher redshift than more traditional line ratio diagnostics because it does not require the use of the \nii/\ha\ ratio.
Using the SDSS DR7 main galaxy sample and the
BPT diagnostic, we calibrate the diagram at low redshift. We find that the diagram can be divided into 3 regions: one occupied mainly
by the pure AGNs (KEx-AGN region), one dominated by composite galaxies
(KEx-composite region), and one contains mostly SFGs (KEx-SFG
region). The new diagram is very efficient for selecting AGNs with high
completeness and low contamination rate.
We further apply the KEx diagram to 7,866 galaxies at 0.3 $<$ z $<$ 1 in the DEEP2 Galaxy Redshift Survey,
and compare the KEx classification to an independent X-ray classification using
$Chandra$ observation. Almost all Type1 AGNs at z$<$0.8 lie in the KEx-AGN region, confirming the reliability of
this classification diagram for emission line galaxies at intermediate redshift.
At z$\sim$2,  the demarcation line between star-forming galaxies and AGNs should be shifted to 0.3~dex higher $\sigma_{\oiii}$ due to evolution

AGNs are separated from SFGs in this diagram mainly because in
addition to that they preferentially reside in luminous and massive
galaxies, they show 0.15/0.12~dex higher $\sigma_{\oiii}$ than
star-forming galaxies at given luminosities/stellar masses. Higher $\sigma_{\oiii}$ also arise from AGN-driven broadening effects (such as winds or outflows). 
When we push to higher redshift, the evolution of \oiii/\hb\ and the TFR
result in the shift of dividing line between AGNs and SFGs. 
KEx needs high enough spectral resolution to measure $\sigma_{\oiii}$, and this diagnostic diagram is 
purely empirical now because it is hard to link ionization and kinematics theoretically. Despite the caveats, 
it provides a robust diagnostic of ionization source when only \oiii\ and \hb\ are available.

\section*{Acknowledgements}
We thank Lisa Kewley, Renbin Yan, Shude Mao, Junqiang Ge, Jong-Hak Woo, Chun Lyu 
for helpful discussions and suggestions. We thank an anonymous referee for helpful suggestions that improve the paper significantly. 
The research presented here is partially supported by the 973 Program of China under grants No. 2013CB834905 and No. 2009CB824800, by the National Natural Science Foundation of China under grants No. 11073040, by the Strategic Priority Research Program  ``The Emergence of Cosmological Structures'' of Chinese Academy of Sciences, Grant No. XDB09000000, and by Shanghai Pujiang Talents Program under grant No. 10pj1411800.
K.Z. acknowledge the supports by the No. 53 China Post-doc general
Fund under grants No. 2013M531232. Funding for the Sloan Digital Sky Survey (SDSS) has been provided by
the Alfred P. Sloan Foundation, the Participating Institutions,
the National Aeronautics and Space Administration, the National Science
Foundation, the U.S. Department of Energy, the Japanese Monbukagakusho,
and the Max Planck Society. The SDSS is managed by the Astrophysical
Research Consortium (ARC) for the Participating Institutions. The SDSS
web site is http://www.sdss.org/.
\newline

\begin{table*}[h]
\topmargin 0.0cm
\evensidemargin = 0mm
\oddsidemargin = 0mm
\scriptsize 
\caption{The statistic of galaxy classification in the KEx diagram}
\label{corrtab}
\medskip
\vfill
\begin{tabular}{l|c c c c | c c c }
\hline \hline
Type             &  Star-Forming Galaxies  & Composites &  LINERs & Seyfert2s & Seyfert1s  & Quasars   &  DEEP2  \\
                   &                                          &                    &                 & (DR7 galaxy sample) & (DR4 quasar sample)  &    \\
(1)              & (2)    & (3)     & (4)      & (5)      &  (6)        & (7)  & (8)     \\
\hline \hline  
Total number     & 97,484 & 16,003  &    998   &   5,860  & 4,624 &   4,158    &   7,866    \\
KEx-SFGs         & 96,322 & 10,416  &    116   &     186  &   414 &      13    &   7,024    \\
KEx-Composite    &    882 &  5,074  &    424   &     335  &   718 &     274    &     344    \\
KEx-AGN          &    280 &    513  &    458   &   5,339  & 3,492 &   3,871    &     498   \\

 \hline
\end{tabular}
\medskip
\vfill
{\normalsize ~Columns from left to right: (1) Classifications on KEx diagram.
 (2)Star-forming galaxies from SDSS main galaxy sample. (3) Composite galaxies from SDSS.
(4) LINERs from SDSS. (5) Seyfert2s from SDSS. (6) Seyfert1s from SDSS main galaxy sample.
(7) Type1 AGNs from Dong et al. (2011).
(8) Galaxies from DEEP2 survey. }\\
\end{table*}

\end{document}